\documentclass[reprint,amsmath,amssymb,aps]{revtex4-2}
\pdfoutput=1

\usepackage{graphicx}
\usepackage{dcolumn}
\usepackage{bm}

\begin{document}

\preprint{APS/123-QED}

\title{Enhanced Imaging of Electronic Hot Spots Using Quantum Squeezed Light}%

\author{Haechan An,$^{1,2}$ Ali Najjar Amiri,$^{2}$ Dominic P. Goronzy,$^{3}$ David A. Garcia Wetten,$^{3}$ Michael J. Bedzyk,$^{3,4}$ Ali Shakouri,$^{1}$ Mark C. Hersam,$^{3,5}$ and Mahdi Hosseini$^{1,2}$}
\email{mh@northwestern.edu}

\address{$^{1}$Elmore Family School of Electrical and Computer Engineering, Purdue University, West Lafayette, Indiana 47907, USA\\
$^{2}$Department of Electrical and Computer Engineering and Applied Physics, Northwestern University, Evanston, Illinois 60208, USA\\
$^{3}$Department of Materials Science and Engineering, Northwestern University, Evanston, Illinois 60208, USA\\
$^{4}$Department of Physics and Astronomy, Northwestern University, Evanston, Illinois 60208, USA\\
$^{5}$Department of Chemistry, Northwestern University, Evanston, Illinois 60208, USA}

\date{\today}

\begin{abstract}
Detecting electronic hot spots is important for understanding the heat dissipation and thermal management of electronic and semiconductor devices. Optical thermoreflective imaging is being used to perform precise temporal and spatial imaging of heat on wires and semiconductor materials. We apply quantum squeezed light to perform thermoreflective imaging on micro-wires, surpassing the shot-noise limit of classical approaches. We obtain a far-field temperature sensing accuracy of 42 mK after 50 ms of averaging and show that a $256\times256$ pixel image can be constructed with such sensitivity in 10 minutes. We can further obtain single-shot temperature sensing of 1.6 K after only 10 $\mathrm{\mu s}$ of averaging enabling dynamical study of heat dissipation. Not only do the quantum images provide accurate spatio-temporal information about heat distribution, but the measure of quantum correlation provides additional information, inaccessible by classical techniques, that can lead to a better understanding of the dynamics. We apply the technique to both Al and Nb microwires and discuss the applications of the technique in studying electron dynamics at low temperatures.
\end{abstract}

\maketitle


Far-field thermal imaging has broad applications ranging from electrical inspection \cite{farzaneh2009ccd,heiderhoff2016thermal,shakouri:opticaexpress2020} to biology \cite{taylor2013biological, Bowen:PhysRevX2014,li2017super,an2020nanodiamond,bai2024single}. Optical detection of hot spots and heat distribution in electronics is particularly important for developing low-power devices and better thermal management. Optical thermoreflective imaging has been used to obtain thermal images with high temporal and spatial resolution in both electronics \cite{shakouri:2010,shakouri:opticaexpress2020} and biological samples \cite{shakouri:scienceadvance2019}. The technique can also be used to study charge transport \cite{karna2023direct} and diffusion \cite{islam2024time} in electronics and superconducting materials. The sensitivity of these classical measurements is fundamentally limited by the shot noise and can be improved by using higher illumination power. However, high-power illumination is not an option when light-induced heating can interfere with measurement or damage the sample. The effect of light-induced heating in low-temperature electronics \cite{olson2019spatially,islam2024time} and biological samples \cite{icha2017phototoxicity} is a well-known issue, and more advancements are needed to overcome the signal-to-noise ratio limit without interfering with the sample's properties.

\begin{figure*}[t] 
\centerline{\includegraphics[width=0.8\textwidth]{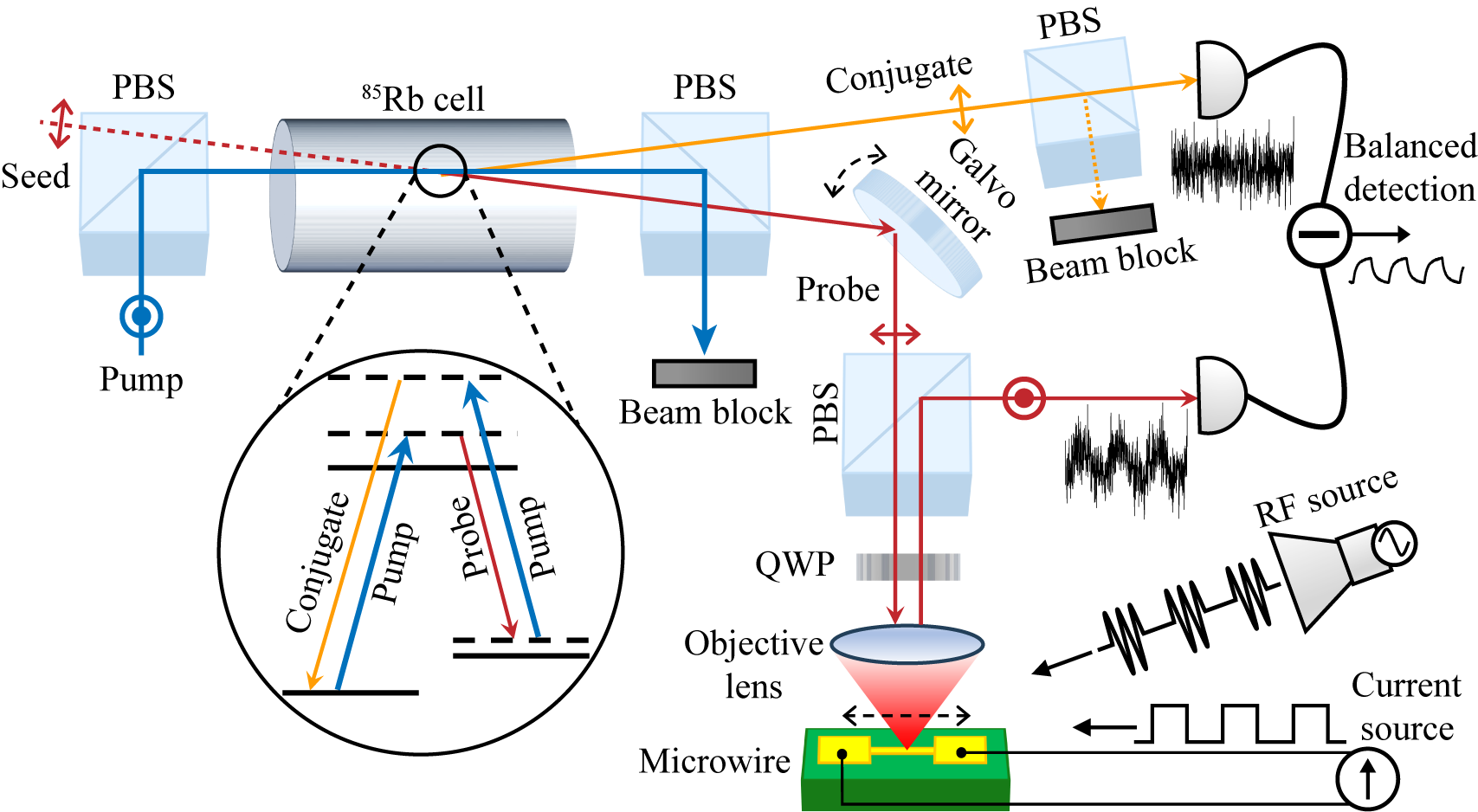}}
\caption{\label{fig1}
Experimental diagram of the quantum photothermal imaging. The four-wave mixing process inside a Rb vapor cell is seeded with a probe light that is amplified at the output leaving the cell together with a correlated conjugate beam at the opposite sides of the pump. The probe is reflected off the microwire and detected after polarization rotation using a quarter-wave plate (QWP) and a polarizing beam splitter (PBS). The difference between the reflected probe and the directly detected conjugate generates a differential signal of reflectivity.}
\end{figure*}

Quantum light is known to break the fundamental limits of classical light and has been proposed in various quantum sensing modalities \cite{Lloyd_sensing:NatPhotonics2018,pooser_lett:sensing2019,Padgett_Imaging:NatRev2019}. In most scenarios, the single-photon-level quantum light has found little real-world applications where the threshold power is much higher than a few photons. Recently, it has been shown that quadrature squeezed states of light can be used to obtain a true quantum advantage when compared to the best classical approaches, in very different contexts, improving the detection of gravitational waves \cite{ligo_nature,ganapathy2023broadband} and single-cell imaging \cite{Bowen:PhysRevX2014,bowen:nature2021,bai2024single}. Bright intensity squeezed states of light \cite{lett:optica2012,Pooser_cantilever:optica2015,wei2019twin,pooser_lett:sensing2019,ather2023quantum,de2024characterizing} are also valuable resources when considering achieving quantum advantage. This is because these optical states can be generated at relatively high powers (milliwatt-level powers) and thus can compete with classical sensing methods applied to certain applications \cite{Vlad_Brillouin:optica2022,ather2023quantum}, provided optical losses are relatively small. Even though applications of squeezed light in various scenarios have been shown \cite{Bowen:PhysRevX2014,pooser_lett:sensing2019,Padgett_Imaging:NatRev2019,prajapati2021quantum}, quantum imaging with squeezed light has not been demonstrated.

In this article, we utilize intensity-squeezed light for imaging heat in microwires. We rely on metallic wires’ non-zero thermoreflective coefficient and sense the temperature by measuring changes in the light’s reflectivity. By scanning the laser spot across a region of the wire, we are able to construct images of hot spots in circuits. The metal’s high reflectivity (or low loss) makes the application of squeezed light particularly relevant to this case. Because the laser-induced heating is comparable to the temperature sensitivity limit of the shot-noise-limited measurement, we conclude that there is a quantum advantage to be gained by using quantum light in this application. To detect temperature's effect on quantum light, we use lock-in amplification and balanced detection. We demonstrate that a two-pixel balanced detection can be used to generate images as fast as a CCD camera.


\begin{figure} 
\centering{\includegraphics{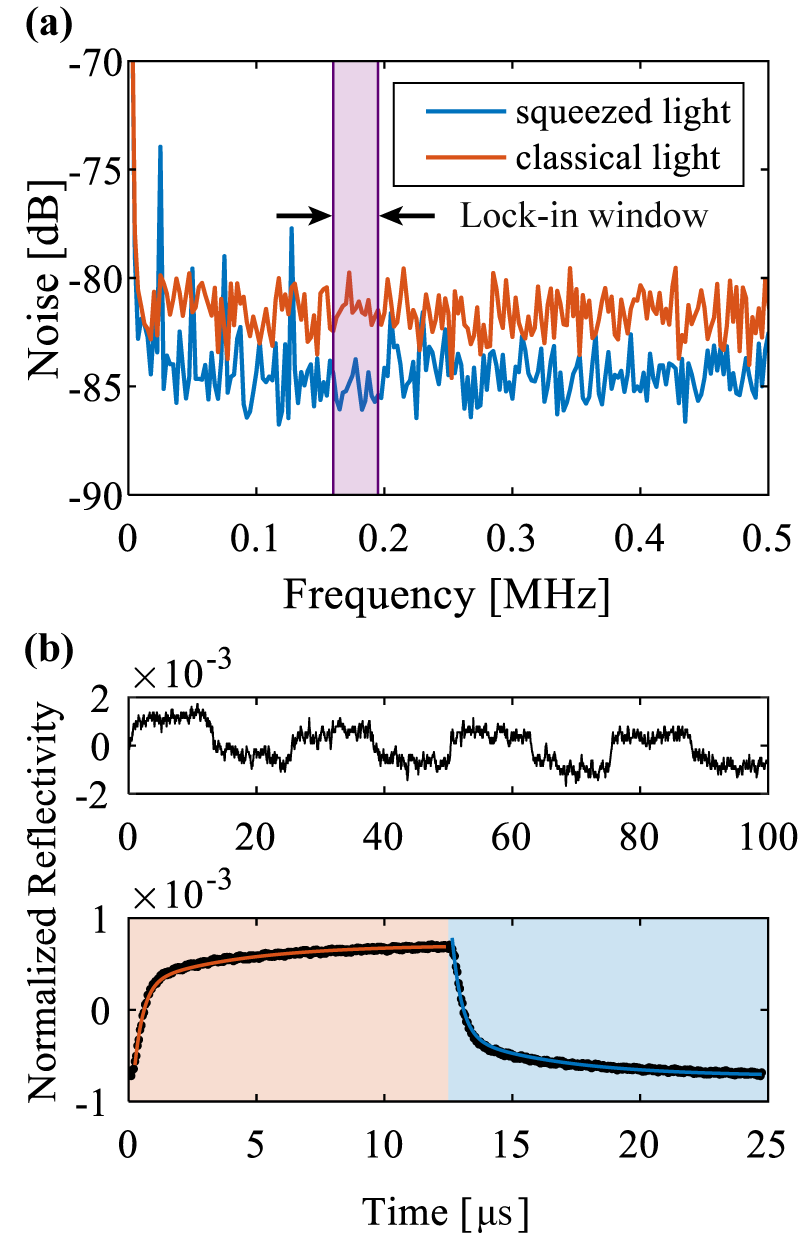}}
\caption{\label{fig2}
(a) Intensity noise measured on a spectrum analyzer for the quantum probe-conjugate pair (blue) as well as a coherent laser beam pair (red) shows intensity squeezing of about 4~dB after reflection compared to the shot-noise limit (measured by the coherent light). (b) The measured reflection for multiple hot and cold frames is shown on the top panel, and the lower panel shows the normalized value of reflectivity averaged over 4000 pulses. The transient behaviors of temperature during heating and cooling cycles are modeled by double exponential functions ($a_0+a_1exp(-t/\tau_1)+a_2exp(-t/\tau_2)$) and shown by solid red and blue lines, respectively.}
\end{figure}

The squeezed light in our experiment is generated via the four-wave mixing (FWM) process in \textsuperscript{85}Rb vapor, similar to our previous experiment \cite{ather2023quantum}. Such FWM process has been used to efficiently generate intensity or quadrature squeezed light \cite{glorieux:PRA2011,corzo2013rotation}.  We use a high-frequency acousto-optic modulator to create a seed probe beam about 3 GHz away from the pump light, both impinging on a one-inch vapor cell heated to 106 $\mathrm{^\circ C}$. Pump and seed probe beams have powers of 600 mW and 5 $\mathrm{\mu W}$, respectively, detuned from the D1 transition by about 1.2 GHz. After the cell, a conjugate beam is generated via the FWM process, whose intensity is correlated to that of the amplified probe beam exiting the cell. The experimental setup is shown in Fig.~\ref{fig1}. The reflected probe light from the wire is detected on a balanced detector and subtracted from the conjugate beam. The probe beam is scanned across an area of the sample, and a change in its reflectivity is recorded as a measure of temperature. The sample’s temperature is modulated by either injecting current into the Al or Nb circuits or remotely coupling an RF signal to Nb waveguides. The Al wire is fabricated using photolithography on silicon dioxide. A thin (5 nm) layer of Ti is deposited on a 300 nm silicon dioxide layer before depositing 100 nm of Al. Photolithography is used to create different circuit geometries of Al wires with a typical wire width of 75 $\mathrm{\mu m}$ and connecting bridges of 10-25 $\mathrm{\mu m}$ wide (acting as hot spots). Coplanar waveguide for the RF circuit is fabricated from a 180 nm Nb film on silicon. For this fabrication, Si(111) was dipped in BOE (5 NH4F : 1 HF) immediately prior to deposition. Deposition was performed using DC magnetron sputtering of a 4N8 purity Nb target at 400 W resulting in a deposition rate of 5 Angstrom/sec. The base pressure of the deposition chamber was less than 1E-9 Torr. The resulting Nb film was patterned via reactive ion etching using SF6.
Nb coplanar waveguide is patterned as 30 $\mathrm{\mu m}$-wide center strip placed between ground plates with 18 $\mathrm{\mu m}$ gaps, giving rise to a characteristic impedance of 50 $\Omega$. The metal's relative reflectivity $\Delta R/R$ is proportional to the metal's temperature change $C_{TR}\Delta T=\Delta R/R$, where $C_{TR}$ is the thermoreflective coefficient. For a shot-noise-limited light, we have a fundamental uncertainity of $\delta(\Delta R/R)=1/\sqrt{N}$, where $N$ is the illumination light's mean photon number. Squeezed light has less noise compared to shot-noise-limited light, and it can provide less uncertainty ($\delta(\Delta R/R)<1/\sqrt{N}$) with the same mean photon number. The active temperature modulation enables lock-in detection to avoid noise near DC and perform measurements in a region of the spectrum where the squeezing is optimum (see Fig.~\ref{fig2} (a)). The temperature modulation of the wire can be controlled by injecting current pulses or RF signals into the wire. Hot (cold) time-frames corresponding to current on (off) give rise to an increase (decrease) in the reflectivity of the probe (see Fig.~\ref{fig2} (b)). The modulation frequency of 40 kHz has optimum squeezing in the frequency range slow enough for wire to reach the quasi-thermal saturation. 

\begin{figure}
\centering
\includegraphics{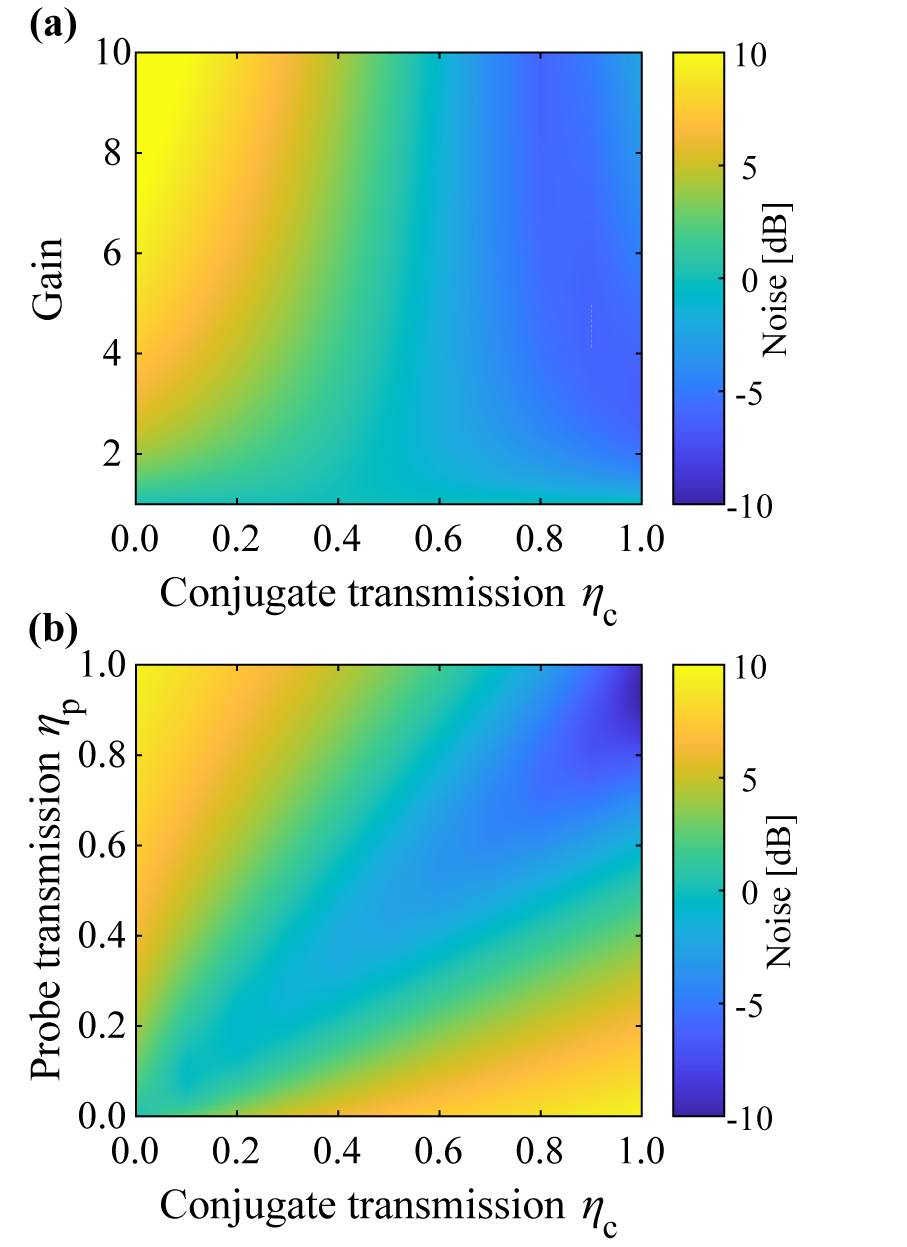}
\caption{\label{fig3}
(a) Calculated squeezing as a function of gain and conjugate loss given a 25 \% probe loss. (b) Calculated squeezing as a function of the probe and conjugate loss given a constant gain of 5.}
\end{figure}

We have a total of 25 \% loss of the probe when going through the imaging system due to the non-unity reflection of the Al wire, scattering, and polarization losses. The amount of squeezing can be optimized given the losses on the probe path by controlling the loss of the conjugate and gain of the probe. The intensity variance is given by the following equation \cite{curvcic2023enhanced}
\begin{equation}\label{eq1}
    V_{loss}=1+2(G-1)\frac{\eta_p^2G-2\eta_p\eta_cG+\eta_c^2(G-1)}{\eta_pG+\eta_c(G-1)}
\end{equation}
where $1-\eta_p$ and $1-\eta_c$ are probe and conjugate loss and $G$ is the FWM gain. Fig.~\ref{fig3} (a) shows the theoretical plot of variance as a function of conjugate loss and gain, given a 25 \% probe loss in our experiment. The optimum squeezing at a specific gain is observed by introducing loss on the conjugate. This is because the probe and conjugate beams individually have noise above shot noise, and sub-shot-noise intensity is only observed after subtracting the beams’ intensities. Therefore, loss on one beam means more uncorrelated noise on the other beam. Fig.~\ref{fig3} (b) depicts variance as a function of the probe and conjugate loss at an FWM gain of 5. It shows that probe and conjugate loss should be balanced to achieve optimum squeezing \cite{curvcic2023enhanced}. We adjust the conjugate loss using a polarizing beam splitter. After optimizing for probe loss in this way, we observe a squeezing of about 4 dB despite the 25 \% probe loss after the cell.


\begin{figure*}
\centering
\includegraphics{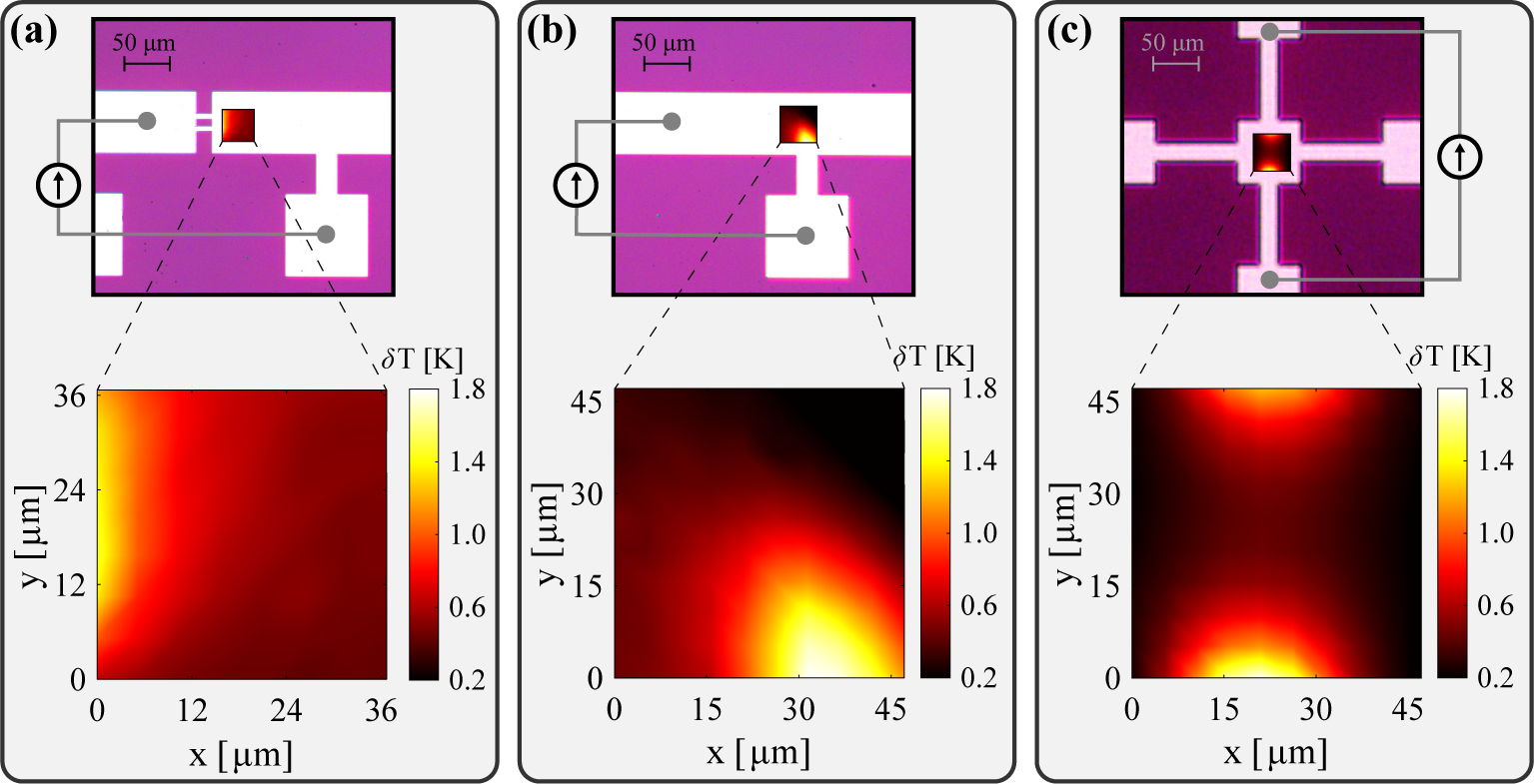}
\caption{\label{fig4}
Microscope images of three different circuits and corresponding thermal maps obtained using quantum thermoreflective imaging on a section of the device (dotted box).}
\end{figure*}

Having optimized squeezing for our specific experiment, we carry out imaging by scanning the probe beam across a region of the wire to map the heat distribution near regions of high current density (expected hot spots). Fig.~\ref{fig4} shows the heat maps for three different circuit geometries obtained using our squeezed light. The beam size is only 25 $\mathrm{\mu m}$ in diameter, limiting the spatial resolution. The spatial resolution can be enhanced simply using a higher NA lens. Each pixel is averaged for 50 ms to obtain a measure of its temperature with an accuracy of 42 mK.  Squeezing of 3-4 dB is maintained during the squeezed light scanning. The 2D Galvo mirror is used to scan the probe beam by 5 $\mathrm{\mu m}$ on the surface of the wire in 0.3 ms, a time duration negligible compared to the single-pixel measurement time. The probe beam center is scanned to create images with step size smaller than the probe's spot size. Fig.~\ref{fig4} (b) \& (c) are constructed from $10 \times 10$ pixels covering an area of 48 $\mathrm{\mu m} \times $48 $\mathrm{\mu m}$, and Fig.~\ref{fig4} (a) is an $8 \times 8$ pixel image. We note that the probe beam has a spot size of 25 $\mathrm{\mu m}$, and near the boundaries of the device, the scattering loss from the tail of the intensity profile limits the imaging area. In the same way, a $256 \times 256$ pixel quantum image with 42 mK temperature resolution can be constructed in 10 minutes. In our case, the probe-induced heating is simulated to be less than 20 mK, which is smaller than the temperature resolution of 42 mK. As a comparison, CCD-based thermal imaging only reaches a temperature sensitivity of 273 mK during the same measurement time. Although it can obtain mega-pixel images, a typical CCD has only a few tens of Hz of frame rate, and this requires a much longer time to achieve high sensitivity. Our imaging method can have a 66 kilo-pixel image with 6 times better sensitivity in the same measurement time. Moreover, by averaging the signal over a single frame (10 $\mathrm{\mu s}$), we can obtain a temperature resolution of 1.6 K. This is a unique advantage of the balanced detector, when compared to most CCD cameras with limited frame rates, that enables tracking fast thermal dynamics in a single measurement. The quantum source is also capable of generating pulsed squeezed light as low as 50 ns long \cite{agha2011time} to further reduce the laser-induced heating effect and perform the lock-in detection at different time windows within a thermal cycle.     

\begin{figure*} 
\centering
\includegraphics{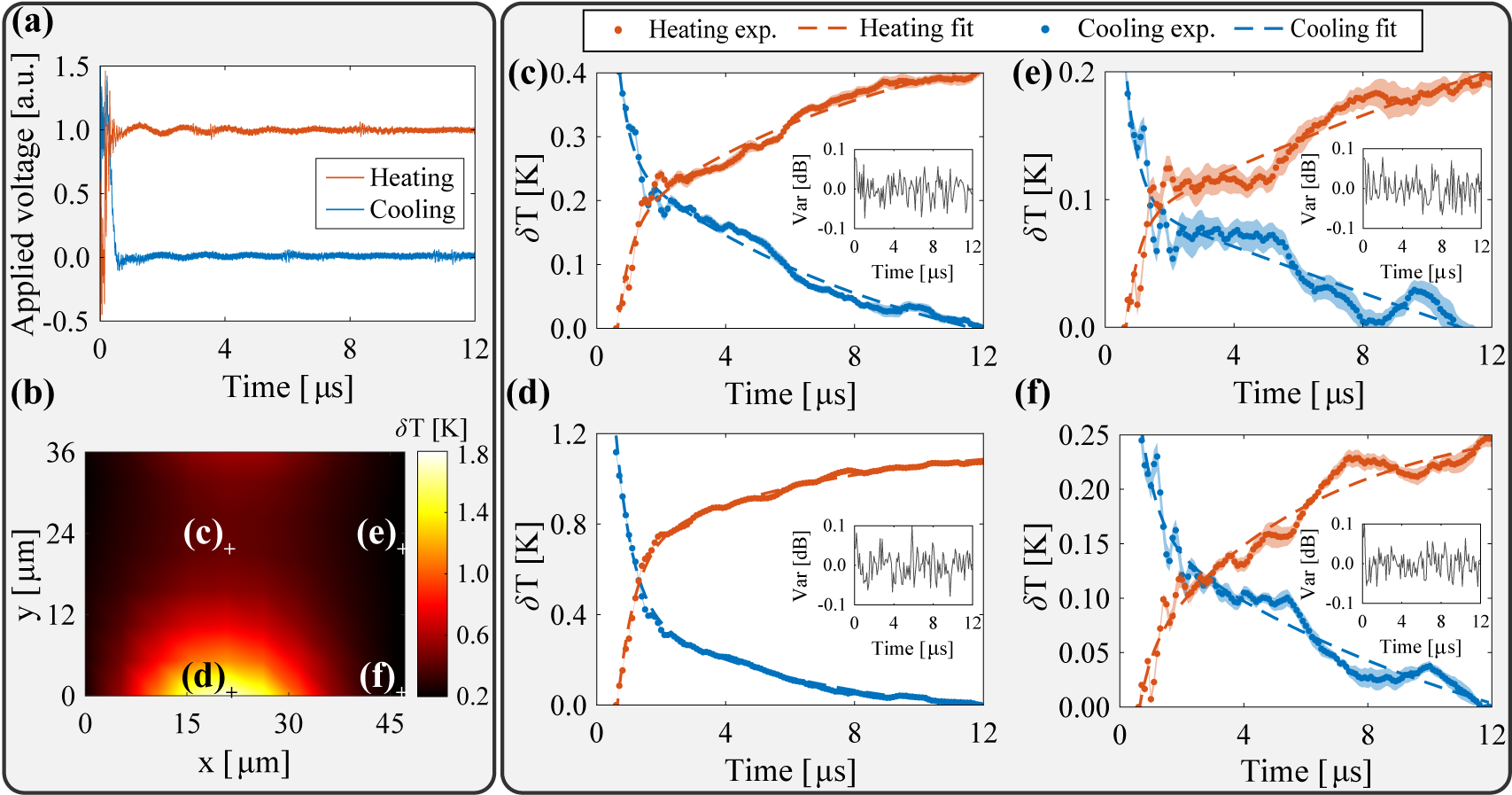}
\caption{\label{fig5}
(a) Applied voltage to the circuit during the heating and cooling cycle. (b) Thermal image obtained from quantum thermoreflevtivity sensing, same as Fig.~\ref{fig4} (c), indicating four regions for which temporal temperature analysis is performed in (c)-(f). (c)-(f) Temperature change data (dot symbols) and its error (shaded area) for the heating (red) and cooling (blue) cycle of four spatial locations in (b). The error bars are obtained by calculating the standard deviation of the temperature change at different time steps. The fitting (shown as dashed lines) is done using double exponential functions ($a_0+a_1exp(-t/\tau_1)+a_2exp(-t/\tau_2)$). Insets in (c)-(f) show the normalized variance calculated from differential intensity of hot and cold frames.}
\end{figure*} 

We can further perform spatial and temporal analysis of the temperature and heat distribution by measuring the reflectivity during different time windows after the current is switched on (heating) or off (cooling) for different spatial locations. Fig.~\ref{fig5} shows the temperature of the device at four different locations with different proximity from the hot spot as a function of time during the heating and cooling cycle. It can be seen that near the hot spot, the temperature is monotonically rising (falling) during the heating (cooling) cycle. However, away from the hot spot, a plateau region is observed due to the competition between heating and cooling processes in the device, namely convection and conduction through different layers. This anomaly is observed in our images thanks to the enhanced temperature resolution. The quantum light used in our images has additional degrees of freedom offering more information about the dynamics of heat dissipation to better understand this anomaly. 
Also plotted in Fig.~\ref{fig5} is the variance of differential intensity of hot and cold frames which represents transient noise response of the measured raw data. The variance remains within 0.1 dB over the cycle period showing consistency of measurement noise during hot and cold frames.
Moreover, the time scale of temperature variation is observed to be different near or away from the hot spot, which is expected to follow a double-exponential curve. The heat transport through the substrate gives rise to a slow dynamic of temperature away from the hot spot. We have confirmed the existence of time scale difference for points close and far from the hot spot using finite-element simulations. As observed from our simulation, the short current pulses and the thin SiO$_2$ layer beneath the Al lead to points away from the hot spot failing to reach thermal equilibrium during the heating-cooling cycle, wherein various heat transport processes compete. This explains the slow temperature rise after 5 $\mathrm{\mu s}$, as seen in Fig.~\ref{fig5} (e) and (f), where a double exponential function is used to fit the data. 

The heat dissipation dynamic has shown anomalous signatures at small scales \cite{ziabari2018full,ku2020imaging}. The high-resolution temperature sensing and negligible laser-induced heating offered by quantum light can help further investigate such dynamics. Akin to fluids, transitioning from laminar flow to turbulence and vortex formation has been predicted and observed in microelectronics \cite{bandurin2016negative,aharon2022direct}. The high temperature resolution provided by quantum light can help further our understanding of fluid-like processes and can aid in designing more efficient electronic devices, relying on measurement-informed thermal management approaches.

\begin{figure*} 
\centering
\includegraphics{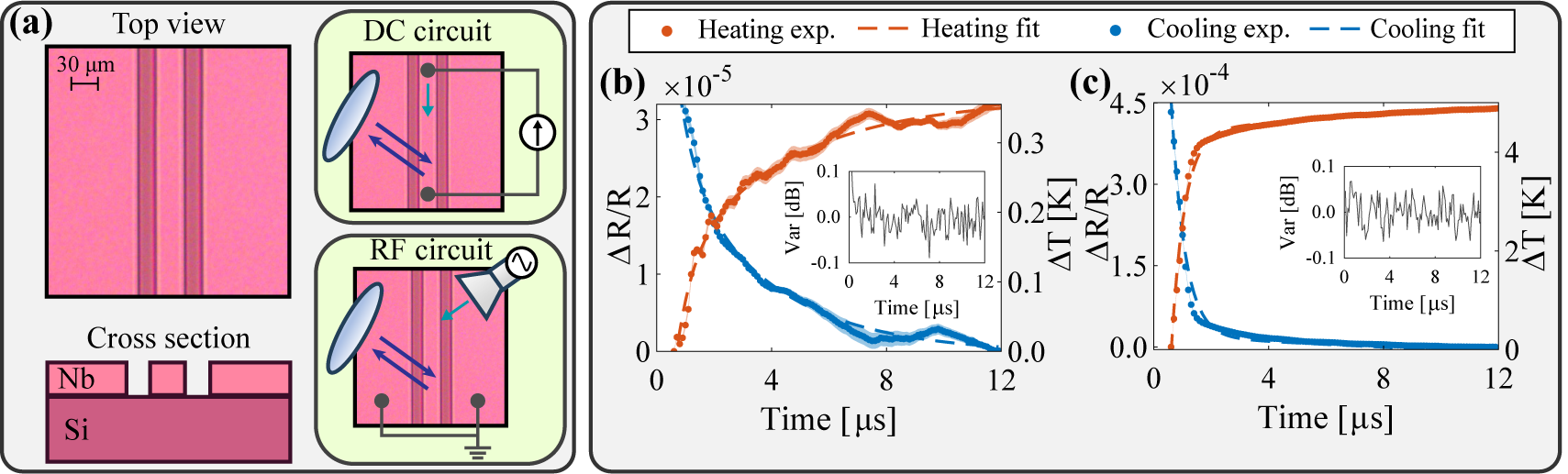}
\caption{\label{fig6}
(a) A top-view microscope image and a schematiccross-section view of the Nb wire structure are shown on the left. Two different configurations, DC and RF, for exciting the Nb wire are also depicted on the right. Thermoreflectivity sensing measurement of a Nb wire of width 30 $\mathrm{\mu m}$ excited by (b) a DC current and (c) a 3.6 GHz RF tone from a horn antenna is shown. The thermoreflective coefficient of the Nb wire is assumed to be $0.9\times10^{-4}$ K \cite{wang2010thermoreflectance}, and the temperature change is calculated accordingly. The change in the reflection of quantum light during the cooling (blue) and heating (red) cycle corresponds to the change in the temperature of the sample, with the error bars shown as shaded areas around the curves. The fitting (shown as dashed lines) is done using double exponential functions ($a_0+a_1exp(-t/\tau_1)+a_2exp(-t/\tau_2)$). Insets in (c)-(f) show the normalized variance calculated from differential intensity of hot and cold frames.}
\end{figure*}

The study of electron-electron and electron-lattice scattering is more relevant for materials with large scattering lengths, where vortex formation in the ballistic or hydrodynamic regimes can be observed. Crystalline Al has an extremely long momentum relaxation length of about 3.3 mm at around 4.2 K temperatures \cite{ueda1996negative}. Carrying out similar measurements as those described above on crystalline Al can be used to study fluid-like dynamics in electronic devices. Another material that has been recently studied for electron vortex formation is WTe$_2$, with large scattering lengths at 4.5 K \cite{aharon2022direct}. Nb is another interesting material used to build superconducting quantum qubits, and understanding defects and charge (or quasi-particle) density fluctuation in these materials can lead to developing better materials and devices with long coherence times \cite{berti2023scanning}. These materials, when measured at low temperatures, are more susceptible to laser heating, and thus quantum sensing can prove to be a valuable tool for studying the physics of these quantum materials, in the far field. To show the feasibility of our quantum sensing technique for Nb, we also measured thermoreflectivity sensing in a Nb coplanar waveguide subject to either current or RF radiation from a horn antenna (see Fig.~\ref{fig6} (a)). Modulation of RF frequency at 3.6 GHz causes modulation of temperature and thus reflectivity (see Fig.~\ref{fig6} (c)). Even though the thermoreflective coefficient of Nb is about 50 \% smaller than Al at 800 nm \cite{wang2010thermoreflectance}, we were able to perform quantum temperature sensing using the same setup for Al wire measurement. The optical loss of Nb is larger than Al, leading to smaller squeezing measured after reflection (1-2 dB). Further improvement of squeezing can be done by adjusting FWM conditions and optimizing the optical gain and loss of our setup. Moreover, the intrinsic multi-spatial-mode nature of squeezed light \cite{boyer2008entangled,swaim2018multimode,an2022intensity,treps2002surpassing}  offers possibility of multimode sensing with cameras \cite{Dowling_PRA2020}.

In conclusion, we utilized quantum intensity-squeezed light to demonstrate the first quantum thermal imaging of hot spots and heat transport in electronics. We demonstrated that using quantum light for hot spot detection provides more details about the dynamics of heat dissipation compared to conventional optical or electronic techniques \cite{vicarelli2012graphene,banadaki2016graphene,christofferson2005thermoreflectance}, thereby enhancing our understanding of thermal processes in electronics. We applied the imaging technique to both Al and Nb circuits with a temperature sensing resolution of 42 mK. The quantum thermoreflective imaging described above can be used to more accurately detect electronic hot spots and heat distribution, study fundamental properties of superconductors by monitoring quasi-particle dynamics \cite{mannila2022superconductor}, hydrodynamic and ballistic effects of electron fluid in metals \cite{aharon2022direct,karna2023direct}, time and frequency domain thermoreflectance to measure thermal conductivity, volumetric heat capacity, thermal transport properties, and electron-phonon coupling in metals \cite{olson2019spatially,islam2024time}.  

\begin{acknowledgments}
We thank Venkat Chandrasekhar for enlightening discussions. We acknowledge funding from the National Science Foundation CAREER Award number: 2144356 and DoD-NDEP Award number HQ0034-21-1-0014. The Nb sample preparation was supported by the U.S. Department of Energy, Office of Science, National Quantum Information Science Research Centers, Superconducting Quantum Materials and System Center (SQMS) under Contract No. DE-AC02-07CH11359. This work made use of the NUFAB Facility, which is supported by the MRSEC program of the National Science Foundation (DMR-2308691) at the Materials Research Center of Northwestern University and the Soft and Hybrid Nanotechnology Experimental Resource (NSF ECCS-2025633).
\end{acknowledgments}

\bibliography{ref}

\end{document}